\documentclass{article}
\usepackage{amsmath}

\input{tcilatex}

\begin{document}

\title{The Hubble parameters in the D-brane models}
\author{Pawel Gusin \\
University of Silesia, Institute of Physics, ul. Uniwersytecka 4, \\
PL-40007 Katowice, Poland,\\
e-mail: pgusin@us.edu.pl }
\maketitle

\begin{abstract}
We consider the DBI action for the D-branes with the dynamic embeddings in
the background produced by p-branes. For the D-brane with the special
topology we obtain two Hublle parameters on this brane. \ The condition for
the equality of these parameters is analyzed. In the special case a mass and
a charge of the background p-branes are derived from this condition.

\begin{description}
\item[PACS] : 11.25.-w ; 11.27.+d

\item[Keywords] : D-branes, p-branes, Hublle parameters
\end{description}
\end{abstract}

\section{Introduction}

The motion of the D-branes in the diversity backgrounds has been considered
in the variety of papers e.g. [1-5]. The applications D-branes to the
cosmology and gravity are also widely discussed e.g. [2-4]. In these
approaches there are considered small size and big size compactificated
directions. The D-brane is considered as an embedded submanifold of the
ten-dimensional spacetime with nontrivial backgrounds fields. The effective
theory is described by the Dirac-Born-Infeld (DBI) action. This action also
determine the motion in the spacetime. From the D-brane point of view this
motion is interpreted as the evolution of the world-volume of the brane. In
the D-brane models of the universe all fields and particles of the Standard
Model (SM) are fixed to the world-volume. Thus evolution of D-brane
corresponds to cosmological evolution for the observer fixed to the
world-volume.

In this paper we consider the cosmological evolution of a probe Dk-brane
with the DBI action in the backgrounds of p-branes. The probe Dk-brane means
that the backreactions are neglected. We assume that the Dk-brane has the
topology of the direct product of a compact space and an non-compact space.
In this case we will obtain the Hubble parameters for these spaces. These
parameters are depend on the tangent and the normal directions to Dk-brane.
The condition for the equality of these parameters is given.

In the section 2 we recall the DBI action for a Dk-brane in the backgrounds
produced by p-branes and derive it equation of motion in given embedding. In
the section 3 we derive a ratio of the Hubble parameters for k=3. They are
obtained from the metric on the world-volume of D3-brane. This case can be
considered as a toy cosmological model corresponding to 3+1 spacetime. In
the general relativity models with several Hubble parameters were considered
long time ago e.g.[6]. In these models space is anisotropic and rate of
expansion depends on directions. From other side cosmological models derived
from M-theory admit warped factors which depend on time [7]. These factors
on IIB string theory side correspond to different rates expansion of the
tangent directions to a D3-brane. The section 4 is devoted to conclusions.

\section{D-brane evolution}

In this section we consider the motion of a Dk-brane. The Dk-brane action is
described by the DBI action:%
\begin{equation}
S=-T_{k}\int d^{k+1}\xi e^{-\phi }\sqrt{-\det (\gamma _{\mu \nu }+2\pi
\alpha ^{\prime }F_{\mu \nu }+B_{\mu \nu })}+T_{k}\int \sum_{i}\widetilde{A}%
_{\left( i\right) }\wedge e^{2\pi \alpha ^{\prime }F+B},  \tag{2.1}
\end{equation}%
where $\gamma _{\mu \nu }$ is the pull back of the background metric, $%
B_{\mu \nu }$ is the pull back of the background the NS 2-form, $F_{\mu \nu
} $ is the strength of the abelian gauge field on the worldvolume and $%
\widetilde{A}_{\left( i\right) }$ are pull-back of the background $i$-forms $%
A_{\left( i\right) }$ with odd (even) degrees: $i=1,3,5,7$ ($i=0,2,4,6,8$)
in the Type IIA (IIB) theory. We consider the background solutions with the
symmetry group $\mathbf{R}^{1}\times E_{\left( 6-p\right) }\times SO\left(
p+3\right) $, where $E_{\left( 6-p\right) }$ is the Euclidean group. They
are given by [8,9,10,11]:

\begin{itemize}
\item the metric:%
\begin{equation}
ds^{2}=-\Delta _{+}\Delta _{-}^{-\frac{7-p}{8}}dt^{2}+\Delta _{+}^{-1}\Delta
_{-}^{\frac{\left( 3-p\right) ^{2}}{2\left( 1+p\right) }-1}dr^{2}+r^{2}%
\Delta _{-}^{\frac{\left( 3-p\right) ^{2}}{2\left( 1+p\right) }}d\Omega
_{p+2}^{2}+\Delta _{-}^{\frac{1+p}{8}}dX_{i}dX^{i},  \tag{2.2}
\end{equation}%
where:%
\begin{equation*}
\Delta _{\pm }\left( r\right) =1-\left( \frac{r_{\pm }}{r}\right) ^{p+1},
\end{equation*}%
$d\Omega _{p+2}^{2}$ is the metric on a (p+2)-dimensional sphere $S^{p+2}$ :%
\begin{equation*}
d\Omega _{p+2}^{2}=h_{rs}d\theta ^{r}d\theta ^{s},
\end{equation*}%
$r,s=1,...,p+2$ and $i=1,...6-p$,

\item the gauge strength $F=dA_{\left( p+1\right) }$ :%
\begin{equation}
F=\left( p+1\right) \left( r_{+}r_{-}\right) ^{\left( p+1\right)
/2}\varepsilon _{p+2},  \tag{2.3 }
\end{equation}%
$\varepsilon _{p+2}$ is the volume form on $S^{p+2}$,

\item the dilaton field:%
\begin{equation}
e^{-2\phi }=\Delta _{-}^{a},  \tag{2.4 }
\end{equation}%
where $a^{2}=\left( 3-p\right) ^{2}/4$.
\end{itemize}

The topological charge $g_{7-p}$ and the mass $m_{7-p}$ of the background
are expressed by $r_{+}$, $r_{-}$ :%
\begin{equation}
g_{7-p}=\frac{vol\left( S^{p+2}\right) }{\sqrt{2}\kappa }\left( p+1\right)
\left( r_{+}r_{-}\right) ^{\left( p+1\right) /2},  \tag{2.5}
\end{equation}%
\begin{equation}
m_{7-p}=\frac{vol\left( S^{p+2}\right) }{2\kappa ^{2}}\left[ \left(
p+2\right) r_{+}^{p+1}-r_{-}^{p+1}\right] .  \tag{2.6}
\end{equation}
The above solution becomes the BPS state for $r_{+}=r_{-}=R$ with the metric:%
\begin{equation}
ds^{2}=\Delta ^{\frac{p+1}{8}}\left( -dt^{2}+dX_{i}dX^{i}\right) +\Delta ^{%
\frac{p-7}{8}}\left( d\rho ^{2}+\rho ^{2}d\Omega _{p+2}^{2}\right) , 
\tag{2.7}
\end{equation}%
where $\rho $ is related to $r$ as follows: $\rho ^{p+1}=r^{p+1}-R^{p+1}$
and $\Delta =1+\left( R/\rho \right) ^{p+1}$.

We have considered this background solution since they are general and for $%
p=3$ the last metric has the form used in the warp compactification.

In the general case the Dk-brane and D(6-p)-brane do not intersect if their
spatial dimensions obey the relation:%
\begin{equation*}
6\geq k+6-p.
\end{equation*}%
We denote the background coordinates as follows:%
\begin{equation*}
X^{M}=(t,X^{1},...,X^{6-p},r,\varphi ^{1},...,\varphi ^{p+2}),
\end{equation*}%
where $\varphi ^{1},...,\varphi ^{p+2}$ are coordinates on the sphere $%
S^{p+2}$, so $r$ and $\varphi ^{1},...,\varphi ^{p+2}$ span the transverse
directions to the (6-p)-brane. The Dk-brane propagating in this background
has $n$-directions parallel to (6-p)-brane and $k-n$ directions
perpendicular to (6-p)-brane where the number $n$ is given by [1]:%
\begin{equation}
n\leq n_{0}=\min (k,6-p).  \tag{2.8}
\end{equation}%
We will consider free falling Dk-brane in its rest frame with the proper
time $\tau $. We assume that $r$ is always transverse to Dk-brane and
Dk-brane has the topology of the direct product:%
\begin{equation*}
V_{n}\times S^{k-n},
\end{equation*}%
where $V_{n}$ is some n-dimensional space. Thus the embedding field has the
form:%
\begin{equation}
X^{M}\left( \tau \right) =\left( t\left( \tau \right) ,\xi ^{1},...,\xi
^{n},X^{n+1},...,X^{6-p},r\left( \tau \right) ,\theta ^{1},...,\theta
^{k-n},\varphi ^{k-n+1}\left( \tau \right) ,...,\varphi ^{p+2}\left( \tau
\right) \right) ,  \tag{2.9}
\end{equation}%
where $\xi ^{1},...,\xi ^{n}$ are coordinates on $V_{n}$ and $\theta
^{1},...,\theta ^{k-n}$ are coordinates on $S^{k-n}$. The induced metric $%
\gamma _{\mu \nu }$ on the world-volume by the embedding (2.9) has the form:%
\begin{equation}
\gamma _{00}=-\Delta _{+}\Delta _{-}^{-\frac{7-p}{8}}\overset{\cdot }{t}%
^{2}+\Delta _{+}^{-1}\Delta _{-}^{\frac{\left( 3-p\right) ^{2}}{2\left(
1+p\right) }-1}\overset{\cdot }{r}^{2}+r^{2}h_{\widehat{r}\widehat{s}}%
\overset{\cdot }{\varphi }^{\widehat{r}}\overset{\cdot }{\varphi }^{\widehat{%
s}},  \tag{2.10}
\end{equation}%
\begin{equation}
\gamma _{ab}=\Delta _{-}^{\frac{1+p}{8}}\delta _{ab},  \tag{2.11}
\end{equation}%
\begin{equation}
\gamma _{\widehat{a}\widehat{b}}=r^{2}h_{\widehat{a}\widehat{b}},  \tag{2.12}
\end{equation}%
where $\widehat{r}$, $\widehat{s}=k-n+1,...,p+2$ , $a,b=1,...n$, and $%
\widehat{a},\widehat{b}=1,...,k-n$. The metrics $h_{\widehat{a}\widehat{b}}$%
\ and $h_{\widehat{r}\widehat{s}}$ are expressed by the metric $h_{rs}$ on
the sphere $S^{p+2}$:%
\begin{equation*}
\left( h_{rs}\right) =\left( 
\begin{array}{cc}
\left( h_{\widehat{a}\widehat{b}}\right) &  \\ 
& \left( h_{\widehat{r}\widehat{s}}\right)%
\end{array}%
\right) .
\end{equation*}%
The dot over coordinates means the derivative with the respect to the proper
time $\tau $. In the case when the background NS form $B$ is zero and the
abelian gauge field on the worldvolume vanishes the WZ term in (2.1) takes
the form:%
\begin{equation*}
\int \widetilde{A}_{\left( k+1\right) },
\end{equation*}%
where the form $\widetilde{A}_{\left( k+1\right) }$ is given by the pull
back of background form $A_{\left( k+1\right) }$. In the considered
background the only non-vanishing form is $A_{\left( p+1\right)
}=A_{M_{0}...M_{p}}dX^{M_{0}}\wedge ...dX^{M_{p}}$ such that $dA_{\left(
p+1\right) }$ is given by (2.3). Thus WZ term does not vanish if k=p and the
DBI action takes the form:

\begin{equation}
S=-T_{k}\int d\tau d^{n}\xi d^{k-n}\theta \left( e^{-\phi }\sqrt{-\det
\left( \gamma _{\mu \nu }\right) }-\delta _{k,p}A\overset{\cdot }{t}\right) 
\tag{2.13}
\end{equation}%
and $A=A_{0...p}$. Since the terms in (2.13) do not depend on coordinates $%
\xi $ we get:%
\begin{equation*}
S=-T_{k}vol\left( V_{n}\right) \int d\tau d^{k-n}\theta \left( e^{-\phi }%
\sqrt{-\det \left( \gamma _{\mu \nu }\right) }-\delta _{k,p}A\overset{\cdot }%
{t}\right) .
\end{equation*}%
\ In the considered background we obtain:%
\begin{gather}
e^{-\phi }\sqrt{-\det \left( \gamma _{\mu \nu }\right) }=\left( \overset{%
\cdot }{t}^{2}-\Delta _{+}^{-2}\Delta _{-}^{\frac{1-p}{1+p}}\overset{\cdot }{%
r}^{2}-\Delta _{+}^{-1}\Delta _{-}^{\frac{7-p}{8}}r^{2}h_{\widehat{r}%
\widehat{s}}\overset{\cdot }{\varphi }^{\widehat{r}}\overset{\cdot }{\varphi 
}^{\widehat{s}}\right) ^{1/2}\times  \notag \\
r^{k-n}\Delta _{+}^{1/2}\Delta _{-}^{[5\left( 1-p\right) +n\left( 1+p\right)
]/16}\sqrt{\det \left( h_{\widehat{a}\widehat{b}}\right) }.  \tag{2.14}
\end{gather}%
In non-rotating case $\overset{\cdot }{\varphi }^{\widehat{r}}=0$ the action
simplifies and takes the form:%
\begin{equation}
S=-T_{k}vol\left( V_{n}\right) \int d\tau L,  \tag{2.15}
\end{equation}%
where the Lagrangian $L$ has the form:%
\begin{equation}
L=\left[ vol\left( S^{k-n}\right) \left( \overset{\cdot }{t}^{2}-\Delta
_{+}^{-2}\Delta _{-}^{\frac{1-p}{1+p}}\overset{\cdot }{r}^{2}\right)
^{1/2}r^{k-n}\Delta _{+}^{1/2}\Delta _{-}^{[5\left( 1-p\right) +n\left(
1+p\right) ]/16}-\delta _{k,p}Aw\overset{\cdot }{t}\right]  \tag{2.16}
\end{equation}%
and $w=\int d^{k-n}\theta $ . Variation $L$ with respect to $t$ gives:%
\begin{equation}
vol\left( S^{k-n}\right) \frac{\overset{\cdot }{t}r^{k-n}\Delta
_{+}^{1/2}\Delta _{-}^{[5\left( 1-p\right) +n\left( 1+p\right) ]/16}}{\sqrt{%
\overset{\cdot }{t}^{2}-\Delta _{+}^{-2}\Delta _{-}^{\frac{1-p}{1+p}}\overset%
{\cdot }{r}^{2}}}-\delta _{k,p}Aw=E,  \tag{2.17}
\end{equation}%
where $E$ is a constant of motion.\ Thus:%
\begin{equation}
\left( \frac{dr}{dt}\right) ^{2}=\left[ 1-\frac{r^{2\left( k-n\right)
}\Delta _{+}^{1/2}\Delta _{-}^{[5\left( 1-p\right) +n\left( 1+p\right) ]/16}%
}{\left( E+\delta _{k,p}Aw\right) ^{2}}vol^{2}\left( S^{k-n}\right) \right]
\Delta _{+}^{2}\Delta _{-}^{-\frac{1-p}{1+p}}.  \tag{2.18}
\end{equation}%
The proper time $\tau $\ of the Dk-brane\ is expressed by:%
\begin{equation*}
d\tau ^{2}=\gamma _{\mu \nu }d\xi ^{\mu }d\xi ^{\nu }=g_{MN}\partial _{\mu
}X^{M}\partial _{\nu }X^{N}d\xi ^{\mu }d\xi ^{\nu }.
\end{equation*}%
In the rest frame of the Dk-brane and for the considering embedding this
proper time has the form:%
\begin{equation}
d\tau ^{2}=-\left( g_{00}+g_{rr}\overset{\cdot }{r}^{2}+r^{2}h_{\widehat{r}%
\widehat{s}}\overset{\cdot }{\varphi }^{\widehat{r}}\overset{\cdot }{\varphi 
}^{\widehat{s}}\right) dt^{2},  \tag{2.19}
\end{equation}%
where:%
\begin{eqnarray*}
g_{00} &=&-\Delta _{+}\Delta _{-}^{-\frac{7-p}{8}}, \\
g_{rr} &=&\Delta _{+}^{-1}\Delta _{-}^{\frac{\left( 3-p\right) ^{2}}{2\left(
1+p\right) }-1}.
\end{eqnarray*}%
In the non-rotating case $\overset{\cdot }{\varphi }^{\widehat{r}}=0$ the
derivatives with the respect to the proper time $\tau $ and coordinate time $%
t$ are related:%
\begin{equation*}
\left( \frac{dr}{dt}\right) ^{2}=\left( \frac{dr}{d\tau }\right) ^{2}\left( 
\frac{d\tau }{dt}\right) ^{2},
\end{equation*}%
so:%
\begin{equation*}
\left( \frac{dr}{dt}\right) ^{2}=-\frac{g_{00}}{1+g_{rr}\left( \frac{dr}{%
d\tau }\right) ^{2}}\left( \frac{dr}{d\tau }\right) ^{2}.
\end{equation*}%
From (2.18) and (2.19) we obtain relation between the radial position and
the proper time:%
\begin{equation}
\left( \frac{dr}{d\tau }\right) ^{2}=\frac{\left( E-\delta _{k,p}Aw\right)
^{2}-r^{k-n}\Delta _{+}^{1/2}\Delta _{-}^{\beta }vol^{2}\left(
S^{k-n}\right) }{\left( E-\delta _{k,p}Aw\right) ^{2}-\Delta _{-}^{\gamma
}+r^{k-n}\Delta _{+}^{3/2}\Delta _{-}^{\delta }vol^{2}\left( S^{k-n}\right) }%
\Delta _{+}\Delta _{-}^{\alpha },  \tag{2.20}
\end{equation}%
where the exponents are:%
\begin{equation*}
\alpha =\frac{-1+14p-p^{2}}{8\left( 1+p\right) },\text{ }\beta =\frac{%
5\left( 1-p\right) +n\left( 1+p\right) }{16},
\end{equation*}%
\begin{equation*}
\gamma =\frac{3\left( p^{2}-10p+9\right) }{8\left( 1+p\right) },\text{ }%
\delta =\frac{p^{2}-60p+61+n\left( 1+p\right) ^{2}}{16\left( 1+p\right) }.
\end{equation*}

In the coordinate time $t$ the induced metric on the Dk-brane by the
embedding (2.9) has the form:%
\begin{equation}
ds^{2}=-\left( \Delta _{+}\Delta _{-}^{-\frac{7-p}{8}}-\Delta
_{+}^{-1}\Delta _{-}^{\frac{\left( 3-p\right) ^{2}}{2\left( 1+p\right) }-1}%
\overset{\cdot }{r}^{2}-r^{2}h_{\widehat{r}\widehat{s}}\overset{\cdot }{%
\varphi }^{\widehat{r}}\overset{\cdot }{\varphi }^{\widehat{s}}\right)
dt^{2}+\Delta _{-}^{\frac{1+p}{8}}d\xi _{a}d\xi ^{a}+r^{2}h_{\widehat{a}%
\widehat{b}}d\theta ^{\widehat{a}}d\theta ^{\widehat{b}}.  \tag{2.21}
\end{equation}%
Using (2.19) we get:%
\begin{equation}
ds^{2}=-d\tau ^{2}+\Delta _{-}^{\frac{1+p}{8}}d\xi _{a}d\xi ^{a}+r^{2}h_{%
\widehat{a}\widehat{b}}d\theta ^{\widehat{a}}d\theta ^{\widehat{b}}, 
\tag{2.22}
\end{equation}%
where $r\left( \tau \right) $ is the solution of the (2.20). This metric has
the form of the FRW-like metric with two scale factors namely $\Delta _{-}^{%
\frac{1+p}{8}}$ and $r^{2}$. Assuming that the evolution of the world-volume
seeing by the observer fixed to the brane, is determined by the gravity
produced by (2.22) we get following equation of motion:%
\begin{equation}
n\left( n-1\right) \overset{\cdot }{\lambda }^{2}+2mn\overset{\cdot }{%
\lambda }\overset{\cdot }{\beta }+m\left( m-1\right) \overset{\cdot }{\beta }%
^{2}+e^{-2\beta }\widetilde{R}=16\pi G\rho ,  \tag{2.23}
\end{equation}%
where $m=k-n$, $\exp 2\lambda =\Delta _{-}^{\frac{1+p}{8}}$, $\exp 2\beta
=r^{2}$, the scalar curvature $\widetilde{R}$ is obtained from the metric $%
h_{\widehat{a}\widehat{b}}$ and $\rho $ is the energy density on the
world-volume. The dot means: $\overset{\cdot }{\lambda }=d\lambda /d\tau $
and so on. Hence from (2.22) follows that the evolution in the non-compact
directions $\xi $ is gven by $\Delta _{-}^{\frac{1+p}{8}}$ and the second
factor $r^{2}h_{\widehat{a}\widehat{b}}$ concerns the evolution in compact
directions corresponding to the sphere.

\section{Hubble parameters}

We relate to the metric (2.22) two Hubble parameters:%
\begin{equation}
H_{n}=\frac{1}{\Delta _{-}^{\frac{1+p}{16}}}\frac{d}{d\tau }\left( \Delta
_{-}^{\frac{1+p}{16}}\right) ,  \tag{3.1}
\end{equation}%
\begin{equation}
H_{c}=\frac{1}{r}\frac{dr}{d\tau },  \tag{3.2}
\end{equation}%
where in $H_{c}$ is assumed isotropic evolution, it means that $d\left( h_{%
\widehat{a}\widehat{b}}\right) /d\tau =0$. The eq. (3.1) takes the form:%
\begin{equation}
H_{n}=\frac{\left( p+1\right) ^{2}}{16}\cdot \frac{rr_{-}^{p+1}}{%
r^{p+1}-r_{-}^{p+1}}\frac{dr}{d\tau },  \tag{3.3}
\end{equation}%
where $dr/d\tau $ is given by (2.20). The ratio of these Hubble parameters
is given by the relation:%
\begin{equation}
\frac{H_{n}}{H_{c}}=\frac{\left( p+1\right) ^{2}}{16}\cdot \frac{%
r^{2}r_{-}^{p+1}}{r^{p+1}-r_{-}^{p+1}}\equiv \eta \left( r\right) . 
\tag{3.4}
\end{equation}%
It depends on the position $r$ of the Dk-brane and $r$ is given by the
solution of the equation (2.20).

We investigate the ratio (3.4) as a function of $t$ in the case when $%
r_{+}=r_{-}=R$. Thus the eq. (2.18) takes the form:%
\begin{equation}
\left( \frac{dr}{dt}\right) ^{2}=\left[ 1-\frac{r^{2\left( k-n\right)
}\Delta ^{1/2+[5\left( 1-p\right) +n\left( 1+p\right) ]/16}}{\left( E+\delta
_{k,p}Aw\right) ^{2}}vol^{2}\left( S^{k-n}\right) \right] \Delta ^{\frac{1+3p%
}{1+p}}  \tag{3.5}
\end{equation}%
and the metric on the world-volume has the form:%
\begin{equation*}
ds^{2}=-d\tau ^{2}+\Delta ^{\frac{1+p}{8}}d\xi _{a}d\xi ^{a}+r^{2}h_{%
\widehat{a}\widehat{b}}d\theta ^{\widehat{a}}d\theta ^{\widehat{b}}.
\end{equation*}%
The ratio of the Hubble parameters in this case is given by:%
\begin{equation}
\frac{H_{n}}{H_{c}}=\frac{\left( p+1\right) ^{2}}{16}\cdot \frac{r^{2}R^{p+1}%
}{r^{p+1}-R^{p+1}}.  \tag{3.6}
\end{equation}%
We restrict ourselves to the case when $k=3\ $which corresponds to D3-brane.
Thus $a,b=1,...n$, and $\widehat{a},\widehat{b}=1,...,3-n$. Then the
eq.(3.5) takes the form:%
\begin{equation}
\left( \frac{dr}{dt}\right) ^{2}=\left[ 1-\frac{r^{2\left( 3-n\right)
}\Delta ^{1/2+[5\left( 1-p\right) +n\left( 1+p\right) ]/16}}{\left( E+\delta
_{3,p}Aw\right) ^{2}}vol^{2}\left( S^{3-n}\right) \right] \Delta ^{\frac{1+3p%
}{1+p}}.  \tag{3.7}
\end{equation}%
In order to get how change (3.6) in time we need find solutions of (3.7).
These solutions among other depends on the dimension $p$ of the background
branes. Thus we have to consider each dimension $p$ separately. The
solutions of (3.7) for differend $p$ are given below where the number of the
non-compact dimensions $n$ is given by the condition (2.8).

For $p=0$ (D-particle) the eq. (3.7) gives following result:%
\begin{equation}
\int \frac{\sqrt{r}dr}{\sqrt{r-R}\sqrt{1-r^{2\left( \alpha -\beta
_{0}\right) }\left( r-R\right) ^{2\beta _{0}}\sigma _{n}^{2}}}=t+t_{0}, 
\tag{3.8}
\end{equation}%
where%
\begin{equation*}
\sigma _{n}^{2}=\left( \frac{vol\left( S^{3-n}\right) }{E}\right) ^{2}
\end{equation*}%
and $\alpha =3-n$ , $2\beta _{0}=\left( 13+n\right) /16$. The number $n$ of
the non-compact dimensions is:%
\begin{equation*}
n=0,1,2,3.
\end{equation*}%
The cases $n=0$ and $n=3$ correspond to the only one Hubble parameter.

For $p=1$ (D-string) we get:%
\begin{equation}
\int \frac{r^{2}dr}{\left( r^{2}-R^{2}\right) \sqrt{1-r^{2\left( \alpha
-\beta _{1}\right) }\left( r^{2}-R^{2}\right) ^{2\beta _{1}}\sigma _{n}^{2}}}%
=t+t_{0},  \tag{3.9}
\end{equation}%
where $2\beta _{1}=\left( 9+2n\right) /16$.

For $p=3$:%
\begin{equation}
\int \frac{r^{5}dr}{\left( r^{4}-R^{4}\right) ^{5/4}\sqrt{1-r^{2\left(
\alpha -\beta _{3}\right) }\frac{\left( r^{4}-R^{4}\right) ^{2\beta _{3}}}{%
\left( E+Aw\right) ^{2}}vol^{2}\left( S^{3-n}\right) }}=t+t_{0},  \tag{3.10}
\end{equation}%
where $2\beta _{3}=\left( 2n-1\right) /8$.

In the both above cases the number $n$ of the non-compact dimensions is
equal to:%
\begin{equation*}
n=0,1,2,3.
\end{equation*}%
The first and the last cases correspond to the only one Hubble parameter.

For $p=5$:%
\begin{equation}
\int \frac{r^{8}dr}{\left( r^{6}-R^{6}\right) ^{4/3}\sqrt{1-r^{2\left(
\alpha -\beta _{5}\right) }\left( r^{6}-R^{6}\right) ^{2\beta _{5}}\sigma
_{n}^{2}}}=t+t_{0},  \tag{3.11}
\end{equation}%
where $2\beta _{5}=3\left( n-2\right) /8$. The number $n$ of the non-compact
dimensions is equal to: $n=0,1$.

The above integrals are complicated. One can evaluated them in the limit
when the parameter $E$ goes to infinity ($E\rightarrow \infty $). In this
case all the above integrals have simply asymptotes: $r\sim t$. It means
that the D3-brane and background $p$-branes does not form bounded system.
Thus one can notice from (3.6) that:%
\begin{equation}
\frac{H_{n}}{H_{c}}=\eta \underset{r\rightarrow \infty }{\rightarrow }%
\left\{ 
\begin{array}{cc}
\infty & p=0 \\ 
R^{2}/4 & p=1 \\ 
0 & p>1%
\end{array}%
\right.  \tag{3.12}
\end{equation}%
and $\eta $ has singularity for all $p$ in $r=R$. As was mentioned above the
considered background solutions are right for $r>R$. So one can conclude
that background produced by D1-branes (D-strings) gives flat Minkowski
4-dimensional space-time with the equal Hubble parameters if $R=2$. This
condition put constraint on a topological charge $g_{6}$ and a mass $m_{6}$
of a dual D6-branes to the background D1-branes, because $R$ is related to
these parameters by (2.5) and (2.6). For $r_{+}=r_{-}=R$ these relations
takes the form:%
\begin{equation}
g_{6}=\frac{3vol\left( S^{3}\right) }{\sqrt{2}\kappa }R^{3},  \tag{3.13}
\end{equation}

\begin{equation}
m_{6}=\frac{3vol\left( S^{3}\right) }{2\kappa ^{2}}R^{3}.  \tag{3.14}
\end{equation}%
Thus we get following values of $g_{6}$ and $m_{6}$:%
\begin{equation}
g_{6}=24\sqrt{2}\pi ^{2}/\kappa ,  \tag{3.15}
\end{equation}%
\begin{equation}
m_{6}=24\pi ^{2}/\kappa ^{2}.  \tag{3.16}
\end{equation}%
In the background produced by D1-branes the condition of the isotropic
expansion leads to the (3.15) and (3.16).

In the case other backgrounds branes one can see from (3.12) that for $p=0$
the expansion of the non-compact dimensions is much faster than compact
dimensions or $H_{c}=0$ which corresponds to the static compact space. For $%
p>2$ the result is that compact dimensions expand faster than non-compact or 
$H_{n}=0$ which gives static non-compact space.

\section{Conclusions}

In this paper we have obtained Hublle parameters for Dk-brane embedded in
the backgrounds produced by the black p-branes. These parameters are related
to the topology of the Dk-brane: the Dk-brane is represented as the
Cartesian product of the $n$-dimensional non-compact space and some $(n-k)$%
-dimensional compact space (in our case this space is sphere). In general
case these parameters have different values. It means that evolution from
the point of view an observer fixed to the Dk-brane in the compact and
non-compact directions is different. The ratio of these parameters has been
obtained in explicit form for big values of $r$. This ratio is equal to one
only in one case for $p=1$ (D-strings) and for special value of $R=2$. It
means that in asymptotic region ($r\rightarrow \infty $) and for $R=2$
expansion is the same in all directions. In this case the mass and the
topological charge are given by eqs. (3.15-3.16 ). The above results are
valid if D3-brane and background branes does not form bounded system. It is
true for sufficient big parameter $E$. In general case the ratio $\eta $
(eq. (3.4)) depends on the position of the D3-brane.

The considered model is an example of a toy cosmological model. The observed
isotropic expansion of our universe is realized in this model as the
condition on equality of Hubble parameters. This condition puts constraint
on the allowed masses and charges of the background D6-branes which are dual
to D1-branes.

Form the other side one can consider this world-volume expansion as driven
by some fictitious fields (mirage cosmology [2]) and postulate that their
energy densities obey the Friedmann equations :

\begin{equation*}
H_{n}^{2}=\rho _{n}\left( \Phi _{n}\right) ,
\end{equation*}%
\begin{equation*}
H_{c}^{2}=\rho _{c}\left( \Phi _{c}\right) ,
\end{equation*}%
where $\rho _{n}\left( \Phi _{n}\right) $\ and $\rho _{c}\left( \Phi
_{c}\right) $\ are the density energy produced by the fictitious scalar
fields $\Phi _{n}$\ and $\Phi _{c}$\ respectively. These fields determinate
cosmological evolution of the world-volume of D3-brane in non-compact and
compact directions.

\section{References}

[1] M. J. Duff, R. R. Khuri, J. X. Lu, \textit{String solitons}, Phys.Rep.
259, 213,(1995), hep-th/9412184 ; M. J. Duff, \textit{Supermembranes},
hep-th/9611203,

[2] A. Kehagis, E. Kiritsis, \textit{Mirage Cosmology},\ hep-th/9910174

[3] E. Kiritsis, \textit{D-branes in Standard Model building, Gravity and
Cosmology}, hep-th/0310001

[4] G. R. Dvali, G. Gabadadze, M. Porrati, Phys. Lett. \textbf{B485} 208
(2000) (hep-th/0005016); G. R. Dvali, G. Gabadadze, Phys. Rev. \textbf{D63 }%
065007 (2001) (hep-th/0008054)

[5] K. L. Panigrahi, \textit{D-brane dynamics in Dp-brane background},
Phys.Lett. B\textbf{601}, 64, (2004), hep-th/0407134

[6] E. Kasner, Am. Journ. Math. \textbf{43}, 217 (1921)

[7] P. Chen, K. Dasgupta, K. Narayan, M. Shmakova, M. Zagermann, \textit{%
Brane Inflation, Solitons and Cosmological Solutions: I}, hep-th/0501185

[8] J. Polchinski, Phys. Rev. Lett. \textbf{75 }(1995) 4724 ; J. Polchinski,
S. Chaudhuri, C. V. Johnson, \textit{Notes on D-Branes}, hep-th/9602052

[9] G. Horowitz, A. Strominger, Nucl. Phys. \textbf{B360}, (1991) 197; M. J.
Duff, J. X. Lu, Nucl. Phys. \textbf{B416 }(1994) 301

[10] M. J. Duff, H. L\"{u}, C. N. Pope, \textit{The Black Branes of M-theory}%
, hep-th/9604052

[11] G. W. Gibons, K. Maeda, \textit{Black Holes and Membranes in Higher
Dimensional Theories with the Dilaton Fields,} Nucl. Phys. \textbf{B207}
(1988) 741; D. Garfinkel, G. T. Horowitz, A. Strominger, \textit{Charged
black holes in string theory},\ Phys. Rev. \textbf{D43 (}1991) 3140-3143;

\end{document}